\documentstyle[12pt]{article}

\def\permil{\%\raise.10ex\hbox{$_{\scriptstyle 0}$}}

\def\beq{\begin{equation}}
\def\eeq{\end{equation}}

\def\bea{\begin{eqnarray}}
\def\eea{\end{eqnarray}}

\begin{document}
\hfill \hspace*{\fill}
\begin{minipage}[t]{3cm}
  DESY-04-104\\
 hep-ph/0406193

\end{minipage}
\vspace*{0.5cm}

\begin{center}
{\Large Solution of the Fan Diagram Equation\\[0pt]
in $2+1$ Dimensional QCD \\[0.5cm]
} {\large \vspace{0.5cm} J. Bartels \\[0pt]
II. Institut f\"{u}r Theoretische Physik, Universit\"{a}t
Hamburg\\[0pt]
Luruper Chaussee 149, D-22761 Hamburg, Germany\\[0.5cm]
V.S.Fadin $^{\dagger }$\\[0pt]
Institute for Nuclear Physics and Novosibirsk State University\\[0pt]
630090 Novosibirsk, Russia\\[0.5cm]
L.N. Lipatov $^{\dagger }$ \\[0pt]
Petersburg Nuclear Physics Institute\\[0pt]
Gatchina, 188 300 St.Petersburg, Russia \\[0.5cm]
}
\end{center}

\vskip15.0pt \centerline{\bf Abstract} 
\noindent 
We investigate the Balitsky-Kovchegov (BK) equation for
$D=3$  space-time dimensions, corresponding to one
transverse coordinate,  and we show that it can be solved
analytically.  The explicit solutions are found in the
linear approximation and for the paricular cases when they
do not depend on the impact parameter of the dipole or on its rapidity.
It is shown, that in a general case different solutions are related by
an infinite parameter group of transformations.
The key observation is that the
equation has the Kowalewskaya-Painlev\'{e} property, which
gives the possibility of reducing the non-linear
problem to the solution of a linear integral equation.

\vskip 0.5cm \hrule \vskip 1cm \noindent \noindent
$^{(\dagger)}$ {\it Humboldt Preistr\"ager
\newline
Work supported in part by INTAS and by the Russian Fund
of Basic Researches} \vfill

\section{Introduction}

In the framework of perturbative QCD the most general
basis for the theoretical description of  small-x
processes is given by the BFKL approach \cite{BFKL},
based on the gluon Reggeization.  Taking into account also
analyticity and unitarity constraints for the $S$-matrix 
one can calculate the high energy asymptotics
of scattering amplitudes with a good accuracy
\cite{LF89}. In particular, using this approach the kernel of
the BFKL equation was obtained
in the next-to-leading order (NLO) \cite{FL98}. To avoid a strong
renormalization scale dependence of its solution in the
$\overline{MS}$-scheme one can chose the
Brodsky-Lepage-Mackenzie (BLM) optimal scale setting
within physical renormalization schemes 
\cite{BFKLP}. It is important that the BLM procedure
allows to preserve, approximately, some good properties of the leading
order approximation, in particular, its conformal invariance \cite{L86}.
Incorporation of the corrections related to 
the renormalization group symmetry \cite{renorm}  should make the 
predictions of the BFKL equation for the small-$x$ physics more
reliable.

However, at asymptotically large energies the BFKL 
approach should be generalized, since it leads to a 
power-like growth of cross sections. The problem of the Froissart
bound violation cannot be solved by the calculation of
radiative corrections at any fixed order and requires
other methods. One of the possibilities is to take into account
the multi-reggeon contributions in the framework of the BKP equations
\cite{BKP}. A general framework for considering 
unitarization effects is the reggeon field theory, 
describing the interactions of reggeized gluons. 
A possible way of deriving such a field theory  
is based on the high-energy effective QCD action \cite{L95},
generalizing the BFKL approach for multi-particle production
and multi-Reggeon interactions.

In a first step, the unitarization program can be realized by
non-linear generalizations of the BFKL equation, related
to the idea of saturation of parton densities
\cite{GLR83}. The most important ingredient is the transition 
vertex for $2 \to 4$ reggeized gluons which, in the BFKL approach, 
has been calculated in \cite{BW} and further investigated in \cite{BLW,BV}.
Using this transition vertex, as a first step beyond the 
BFKL equation, one can formulate a nonlinear evolution equation which sums 
the QCD fan diagrams consisting of BFKL Green's functions ~\cite{BLV}.     
In a future step towards finding solutions to the reggeon field theory, 
closed loops of BFKL Pomerons should be included and calculated.   
  
At present it is fashionable to describe the
saturation in DIS  within the color dipole approach
\cite{dipole}, which in the target rest frame has an attractive and clear 
physical interpretation: the incoming $\gamma ^{\ast }$
splits into a $q\bar{q}$ colour dipole long before the
dipole interacts with the target. In this approach the
cross section  is presented as
\begin{equation}
\sigma _{\gamma ^{\ast }}(x,Q^{2})=\int
d^{2}r\int_{0}^{1}dz|\Psi _{\gamma ^{\ast
}}(r,z,Q^{2})|^{2}{\sigma }_{dp}(r,x),
\label{sigmagamma}
\end{equation}
where $\Psi _{\gamma ^{\ast }}(r,z,Q^{2})$ is the photon
wave function, $z$ the quark momentum fraction, and $r =
|\vec r|$ the transverse size of the colour dipole,
$\vec{r}=\vec{r}_{1}-\vec{r}_{2}$; the transverse vectors
$\vec{r}_{1}$ and $\vec{r}_{2}$ are the quark and
antiquark coordinates, and ${\sigma }_{dp}(r,x)$ is
the dipole cross section,
\begin{equation}
\sigma _{dp}(r,x)=2\int d^{2}b\left(
1-S(\vec{r}_{1},\vec{r}_{2};Y)\right) \,.
\label{sigmadp}
\end{equation}
Here $\vec{b}=(\vec{r}_{1}+\vec{r}_{2})/2$ is the impact
parameter of the quark-antiquark system,  $Y=\log (1/x)$,
and $\,S(\vec{r}_{1},\vec{r}_{2};Y)= \langle tr\left( U^{+}(\vec{r}
_{1})U(\vec{r}_{2})\right) /N_{c}\rangle _{Y}$ is the
$S$-matrix for the elastic scattering , which can be
presented as an average of the product of two Wilson
lines\\ 
$U^{+}(x_{\perp })=P\exp [ig\int dx^{-}A_{a}^{+}(x^{-},x_{\perp })t^{a}]\,.$

There exist two different approaches for  calculating
the $Y$ dependence of $\,S(%
\vec{r}_{1},\vec{r}_{2};Y)$. The first one starts from an
infinite hierarchy of coupled equations \cite{Balitsky}
for products of any given number of Wilson lines. Taking
the limit of large $N_{c}$ and considering the case of
the target being a large nucleus, the equation for the
two-line correlator decouples and takes a simple form
\cite{Kovchegov}:
\[
\frac{\partial S(\vec{r}_{1},\vec{r}_{2},Y)}{\partial
Y}=\frac{\alpha
_{s}N_{c}}{2\pi ^{2}}\int d^{2}\vec{r}\frac{(\vec{r}_{1}-\vec{r}_{2})^{2}}{(%
\vec{r}_{1}-\vec{r})^{2}(\vec{r}_{2}-\vec{r})^{2}}\left[ S(\vec{r}_{1},\vec{r%
},Y)S(\vec{r},\vec{r}_{2},Y)\right.
\]
\begin{equation}
\left. \ -S(\vec{r}_{1},\vec{r}_{2},Y)\right] \,.
\label{BK}
\end{equation}
At small $N=1-S$, one can neglect terms nonlinear in $N$,
and this equation turns into the colour-dipole version of
the BFKL equation for the dipole cross section $N$. In
\cite {BLV} it has been shown that Eq.(\ref{BK}) can be
considered as a special limit of the QCD fan diagram equation mentioned 
before: apart from taking the large-$N_c$ limit, the dipole cross section 
has to be in the M\"obius class of functions. An
alternative approach for describing saturation is the
Color Glass Condensate \cite{CGC}. In this framework, the
$Y$-evolution is based upon functional equations. It is
affirmed \cite{Blaizot} that both approaches give
identical results for the calculation of observables.

Eq.(\ref{BK}) has attracted much interest. First,
numerical solutions have been found for a $b$-independent
approximation \cite{num1}, and later on also for the full
$b$-dependent equation \cite{num2}. Several attempts have
recently been made to find analytical solutions, at least
in some approximation \cite{approx}. However, so far no
exact analytical solution is known. Therefore it will be
interesting to find a model in which the BK equation
could be solved.

In this paper we generalize the BK-equation (\ref{BK}) to
an arbitrary dimension $D$ of the space-time. For the
case $D=2+1$ - which corresponds to one transverse
dimension - we shall present analytical solutions (the
BFKL dynamics was investigated at $D=2+1$ in the papers
\cite{Tan}). The key observation in finding  these
solutions is that the equation has the
Kowalewskaya-Painlev\'{e} property. We proceed in
several steps of approximations. After writing down the
equation in one transverse dimension (Section 2) we first
discuss the linear approximation (Section 3). Then we
turn to the nonlinear equation, beginning with the
$b$-independent case (Section 4), where we find explicit
solutions for arbitrary initial conditions. Passing on to
the $b$-dependence we first find all "stationary"
(rapidity independent) solutions (Section 5),
generalizing then to the rapidity dependent case (Section
6). We conclude with  the investigation of the infinite parameter 
symmetry of the equation
(Section 7) and its general solution (Section 8). A few
details are put into the Appendix.

\section{The fan equation for 2+1 QCD}

Let us begin by writing down the fan equation for the
$S$-matrix describing the dipole evolution in the
($2+1$)-dimensional QCD. The kernel of the BK equation
(\ref{BK}) is equal to the probability density in the
space of transverse coordinates and rapidity  for the
soft gluon emission by the quark-antiquark pair:
\begin{equation}
\frac{\alpha _{s}N_{c}}{2\pi ^{2}}\frac{(\vec{r}_{1}-\vec{r}%
_{2})^{2}}{(\vec{r}_{1}-\vec{r})^{2}(\vec{r}_{2}-\vec{r})^{2}}%
= 2 \alpha_{s}N_{c}\mid\int \frac{d^{D-2}k}{(2\pi)^{D-2}}
\frac{\vec k}{\vec{k}^{2}}(e^{i\vec{k}(\vec r_1-\vec
r)}-e^{i\vec{k}(\vec r_2-\vec r)})\mid ^{~2}\,
\end{equation}
at $D=4$.  It is clear that if the space-time dimension
takes the value $D=4+2\epsilon$, the BK equation
preserves its form, but since
\begin{equation}
\int \frac{d^{2+2\epsilon }k}{(2\pi )^{2+2\epsilon }}\frac{\exp {(i\vec{k}%
\vec{r})}\vec{k}}{\vec{k}^{2}}=\frac{i}{2\pi}\frac{\Gamma
(1+\epsilon )}{\pi ^{\epsilon }}\left( \frac{\vec{r}}{(\vec{r}%
^{2})^{1+\epsilon }}\right), 
\end{equation}
the kernel changes according to
\begin{equation}
\frac{\alpha _{s}N_{c}}{2\pi ^{2}}\frac{(\vec{r}_{1}-\vec{r}%
_{2})^{2}}{(\vec{r}_{1}-\vec{r})^{2}(\vec{r}_{2}-\vec{r})^{2}}%
\longrightarrow \frac{\alpha _{s}N_{c}}{2\pi ^{2}}\left(
\frac{\Gamma
(1+\epsilon )}{\pi ^{\epsilon }}\right) ^{2}\left(
\frac{(\vec{r}_{1}-\vec{r})}{(\vec{r}_{1}-\vec{%
r})^{2(1+\epsilon )}}-\frac{(\vec{r}_{2}-\vec{r})}{(\vec{r}_{2}-\vec{r}%
)^{2(1+\epsilon )}}\right) ^{2}\,.
\end{equation}
Putting $\epsilon = -1/2$ we obtain:
\begin{equation}
\frac{\partial S_{\rho _{2}\rho _{1}}(y)}{\partial
y}=\int_{\rho _{1}}^{\rho _{2}}d\rho _{0}\left( S_{\rho
_{2}\rho _{0}}(y)S_{\rho _{0}\rho _{1}}(y)-S_{\rho
_{2}\rho _{1}}(y)\right) \,,\,y=\frac{\alpha_s}{2\pi }Y~,
\label{fan}
\end{equation}
where the inequality $\rho _{2}>\rho _{1}$ is imposed
without any restriction of generality.  Note that the
scattering amplitude $N_{\rho _{2}\rho _{1}}(y)$ is
related to the $S$-matrix $S_{\rho _{2}\rho _{1}}(y)$ by
the identity
\begin{equation}
N_{\rho _{2}\rho _{1}}(y)=1-S_{\rho _{2}\rho _{1}}(y)\,.
\end{equation}
We introduce the dipole impact parameter $b$ and its size
$\rho $
\begin{equation}
b=\frac{\rho _{2}+\rho _{1}}{2}\,,\,\,\rho =\rho
_{2}-\rho _{1}\,,\,\,\rho ^{\prime }=\rho _{0}-\rho
_{1}~.
\end{equation}
Then the fan equation can be presented as follows
\begin{equation}
\frac{\partial S(b,\rho ,y)}{\partial y}=\int_{0}^{\rho
}d\rho ^{\prime
}\left( S(b+\frac{\rho ^{\prime }}{2},\rho -\rho ^{\prime },y)\,S(b-\frac{%
\rho -\rho ^{\prime }}{2},\rho ^{\prime },y)-S(b,\rho
,y)\right)\,, \label{fan1}
\end{equation}
or, in a simpler way,
\begin{equation}
\left( \frac{\partial }{\partial y}+\rho \right)
\,S(b,\rho ,y)=\int_{0}^{\rho }d\rho ^{\prime
}S(b-\frac{\rho ^{\prime }}{2},\rho -\rho ^{\prime
},y)\,S(b+\frac{\rho -\rho ^{\prime }}{2},\rho ^{\prime
},y)\,. \label{fan2}
\end{equation}

Now we go to the mixed representation ($b,p$)
\begin{equation}
S(b,\rho ,y)\,\theta (\rho )=\int_{-i\infty }^{i\infty }\frac{dp}{2\pi i}%
\,e^{\rho p}\,s(b,p,y)\,
\end{equation}
with the Laplace transform
\begin{equation}
s(b,p,y)\,=\int_{0}^{\infty }d\rho \,e^{-\rho
p}\,S(b,\rho ,y).
\end{equation}
In this representation $\rho $ is the differential
operator
\begin{equation}
\rho \,s(b,p,y)=-\frac{\partial }{\partial
p}\,s(b,p,y)\equiv -\partial s(b,p,y)\,.
\end{equation}

The convolution of two functions in the $\rho
$-representation
\begin{equation}
F_{12}(\rho )=\int_{0}^{\rho }d\rho ^{\prime }F_{1}(\rho
-\rho ^{\prime })\,\,F_{2}(\rho ^{\prime })\,
\end{equation}
is reduced to the product of their Mellin transforms
\begin{equation}
f_{12}(p)=f_{1}(p)\,\,f_{2}(p)\,,\,\,\,f(p)\,=\int_{0}^{\infty
}d\rho \,e^{-\rho p}\,F(\rho )\,.
\end{equation}
Moreover,  a more general relation holds
\begin{equation}
f_{12}^{kl}(p)=\left( \partial ^{l}f_{1}(p)\right)
\,\,\left( \partial ^{k}f_{2}(p)\right) \,\,\,\,
\end{equation}
for the Mellin transform of the function
\begin{equation}
F_{12}^{kl}(\rho )=\int_{0}^{\rho }d\rho ^{\prime
}\,(-\rho ^{\prime })^{k}(-\rho +\rho ^{\prime
})^{l}F_{1}(\rho -\rho ^{\prime })\,F_{2}(\rho ^{\prime
})\,.
\end{equation}
Let us introduce the anti-normal ordering $\overline{N}$
of operators depending on the two momenta $p_{r}$
($r=1,2$) and the corresponding derivatives $\partial
_{r}=\partial /\partial (p_{r})$, according to the rules
\begin{equation}
\overline{N}(p_{r}\partial _{s})=\overline{N}(\partial
_{s}p_{r})\equiv
\partial _{s}p_{r}=p_{r}\partial _{s}+\delta _{rs}\,.
\end{equation}
Then one can verify that the fan equation in the mixed representation ($b,p$%
) can be written in the form
\begin{equation}
\left(\frac{\partial }{\partial y }-\frac{\partial
}{\partial p }
\right)s(b,p,y)=\lim_{p_{r}\rightarrow p}\overline{%
N}\,\left( s(b+\frac{1}{2}\partial
_{2},p_{1},y)\,s(b-\frac{1}{2}\partial
_{1},p_{2},y)\right) 1\,, \label{equation in mixed
representation}
\end{equation}
where it is implied that, after performing the
transformation of the differential operator to the
anti-normal form, one omits all terms with the
differential operators due to the relations
\[
\partial _{1}^{l}\partial _{2}^{k}1=0\,,\,\,k,l\geq 1
\]
and, at the end, puts $p_{1}=p_{2}=p$.

Note that one can consider a modified equation for the
$S$-matrix in the case of a non-zero temperature $T$ in
the $t$-channel (see \cite{pomterm}). In this case
$S(\rho ,y)$ is assumed to be a periodic function of
$\rho _{k}$
\begin{equation}
S_{\rho _{2}\rho _{1}}(y)=S_{\rho _{2}\rho
_{1}}(y)_{|\rho _{k}\rightarrow \rho _{k}+1/T}\,.
\end{equation}

\section{The BFKL equation in $2+1$ dimensional QCD}

To begin with, we consider the linearized case
corresponding to
the BFKL equation in $2+1$ dimensional space-time. Using the relation $%
S(b,\rho ,y)=1-N(b,\rho ,y)$ and neglecting nonlinear
contributions in (\ref {fan2}), one can obtain
\begin{equation}
\left( \frac{\partial }{\partial y}+\rho \right)
\,N(b,\rho
,y)=\int_{0}^{\rho }d\rho ^{\prime }\left( N(b-\frac{\rho ^{\prime }}{2}%
,\rho -\rho ^{\prime },y)+N(b+\frac{\rho ^{\prime
}}{2},\rho -\rho ^{\prime },y)\right) \,. \label{linear
fan}
\end{equation}
This equation is simplified significantly in the case
when $N$ does not depend on $b$
\begin{equation}
\left( \frac{\partial }{\partial y}+\rho \right) \,N(\rho
,y)=2\int_{0}^{\rho }d\rho ^{\prime }N(\rho ^{\prime
},y)\,.\label{linear b-independent fan}
\end{equation}
For the Mellin transform
\begin{equation}
n(p,y)\,=\int_{0}^{\infty }d\rho \,e^{-\rho p}\,N(\rho
,y)\,,\;\;\;N(\rho ,y)\,\theta (\rho )=\int_{-i\infty
}^{i\infty }\frac{dp}{2\pi i}\,e^{\rho p}\,n(p,y)
\end{equation}
we obtain the differential equation
\begin{equation}
(\frac{\partial }{\partial y}-\frac{\partial }{\partial p})\,n(p,y)=\frac{2}{%
p}\,n(p,y)\,.\label{differential b-independent fan}
\end{equation}
Its solution with the initial condition $n(p,0)=n_{0}(p)$
is
\begin{equation}
n(p,y)=\frac{(y+p)^{2}\,n_{0}(y+p)}{p^{2}}\,.
\end{equation}
The second order pole at $p=0$ provides a linear
dependence of $N(\rho ,y)$
at large $\rho $%
\begin{equation}
N(\rho ,y)\simeq \rho \,y^{2}\,n_{0}(y)\,,
\end{equation}
which leads to a violation of the $S$-matrix unitarity at
$y\neq 0$ for any initial condition. Note that
Eq.(\ref{linear b-independent fan}) has a stationary
($y$-independent) solution $N(\rho ,y)=c\rho $, with
arbitrary $c$. But this solution is not acceptable from
the physical point of view.

Let us turn now to Eq.(\ref{linear fan}) for the
amplitude depending on $b$. Performing the Mellin and
Fourier transformations in the variables $\rho $ and $b$,
respectively,
\[
N(b,\rho ,y)\,\theta (\rho )=\int_{-i\infty }^{i\infty }\frac{dp}{2\pi i}%
\,e^{\rho p}\int_{-\infty }^{\infty }\frac{dq}{2\pi
}\,e^{iqb}\,n(q,p,y)\,,
\]
\begin{equation}
n(q,p,y)\,=\int_{0}^{\infty }d\rho \,e^{-\rho p}\int_{-\infty }^{\infty }{db}%
\,e^{-ibq}\,N(b,\rho ,y), \label{linear fan mellin})
\end{equation}
we obtain
\begin{equation}
(\frac{\partial }{\partial y}-\frac{\partial }{\partial p})\,n(q,p,y)=\frac{%
8p}{4p^{2}+q^{2}}n(q,p,y)\,. \label{differential
b-dependent BFKL}
\end{equation}
Its general solution is
\begin{equation}
n(q,p,y)=\frac{f(p+y;q)}{4p^{2}+q^{2}}\,, \label{solution
of differential b-dependent BFKL}
\end{equation}
with the arbitrary function $f(p+y;q)$. For the initial condition $%
n(q,p,0)=n_{0}(q,p)$ the result can be written as follows
\begin{equation}
n(q,p,y)=\frac{(4(y+p)^{2}+q^{2})\,n_{0}(q,y+p)}{4p^{2}+q^{2}}\,,
\end{equation}
where
\begin{equation}
n_{0}(q,p)=\int_{0}^{\infty }d\rho \,e^{-\rho p}\int_{-\infty }^{\infty }{db}%
\,e^{-ibq}\,N(b,\rho ,0)\,.
\end{equation}
Therefore
\[
N(b,\rho ,y)\,=\int_{-i\infty }^{i\infty }\frac{dp}{2\pi
i}\,e^{\rho
p}\int_{-\infty }^{\infty }\frac{dq}{2\pi }\,e^{iqb}\,\frac{%
(4(y+p)^{2}+q^{2})\,}{4p^{2}+q^{2}}
\]
\begin{equation} \times \int_{0}^{\infty }d\rho ^{\prime
}\,e^{-\rho ^{\prime }(p+y)}\int_{-\infty }^{\infty
}{db}^{\prime }\,e^{-ib^{\prime }q}\,N(b^{\prime },\rho
^{\prime },0)\,.\label{solution of linear fan}
\end{equation}
By introducing the Green function $G_{y}(b,\rho
;b^{\prime },\rho ^{\prime },0)$ according to the
definition
\begin{equation}
N(b,\rho ,y)=\int_{0}^{\infty }d\rho ^{\prime }\,e^{-\rho
^{\prime }y}\int_{-\infty }^{\infty }{db}^{\prime
}\,G_{y}(b,\rho ;b^{\prime },\rho ^{\prime
})\,N(b^{\prime },\rho ^{\prime },0)\,,
\end{equation}
we obtain for it
\[
\,G_{y}(b,\rho ;b^{\prime },\rho ^{\prime })\,= \delta
(b-b^{\prime })\delta (\rho -\rho ^{\prime })
\]
\begin{equation}
+ \theta (\rho -\rho ^{\prime })\int_{-\infty }^{\infty
}\frac{dq}{2\pi }\,e^{iq(b-b^{\prime })}\,\left(2y\cos
\frac{q(\rho -\rho ^{\prime })}{2}+ \frac{2y^{2}}{q}\sin
\frac{q(\rho -\rho ^{\prime })}{2}\right)\,,
\end{equation}
so that the Green function for the BFKL Pomeron is a
polynomial in $y$
\[
G_{y}(b,\rho ;b^{\prime },\rho ^{\prime })\,=\delta
(b-b^{\prime })\,\delta (\rho -\rho ^{\prime })
\]
\begin{equation}
+\theta (\rho -\rho ^{\prime })\sum_{\lambda =\pm
}\left( y\,\delta (b-b^{\prime }+\lambda \frac{\rho -\rho ^{\prime }}{2}%
)+\lambda y^{2}\theta (b-b^{\prime }+\lambda \frac{\rho -\rho ^{\prime }}{2}%
)\right) .
\end{equation}
Therefore the $y$-evolution of the initial dipole
distribution $N_{\rho _{2}\rho _{1}}(0)$ in \ the\ ($\rho
_{2},\rho _{1}$)-representation is simple:
\[
N_{\rho _{2}\rho _{1}}(y)=e^{-y\rho _{21}}N_{\rho
_{2}\rho _{1}}(0)+y\,\int_{\rho _{1}}^{\rho _{2}}d\rho
_{0}\left( e^{-y\rho _{20}}N_{\rho _{2}\rho
_{0}}(0)+e^{-y\rho _{01}}N_{\rho _{0}\rho _{1}}(0)\right)
\]
\begin{equation}
+y^{2}\int_{\rho _{1}}^{\rho _{2}}d\rho _{2}^{\prime
}\int_{\rho _{1}}^{\rho _{2}^{\prime }}d\rho _{1}^{\prime
}e^{-(\rho _{2}^{\prime }-\rho _{1}^{\prime })y}N_{\rho
_{2}^{\prime }\rho _{1}^{\prime }}(0).
\end{equation}

For large $y\gg 1/\rho $, where in (\ref{solution of differential
b-dependent BFKL}) one can neglect $p$ in
comparison with $y$, the solution in the coordinate
representation is
\begin{equation}
N(b,\rho ,y)\,=\int_{-\infty }^{\infty }\frac{dq}{2\pi }\,e^{iqb}\,\frac{%
\sin (\rho q/2)}{2q}(4y^{2}+q^{2})\,n_{0}(q,y)\,.
\end{equation}
In particular, for large $b\gg \rho$ we obtain the above
discussed expression $N(b,\rho ,y)\simeq \rho
\,y^{2}\,n_{0}(y)$. On the other hand, at $b\ll \rho$ one
has
\begin{equation}
N(b,\rho ,y)\,\simeq \int_{-\infty }^{\infty
}\frac{dq}{2\pi }\,\,\frac{\sin (\rho
q/2)}{q}\,2y^{2}\,n_{0}(q,y)\,.
\end{equation}

Note, that Eq. (\ref{differential b-dependent BFKL})
admits stationary solutions
\begin{equation}
n(q,p,y)\,=\frac{f(q)}{4p^{2}+q^{2}}\,
\end{equation}
with an arbitrary function $f(q)$. For $N(b,\rho ,y)$ we
obtain correspondingly
\begin{equation}
N(b,\rho ,y)\,=a(b+\frac{\rho }{2})-a(b-\frac{\rho
}{2})\, \label{stationary N}
\end{equation}
with an arbitrary amplitude $a$. Actually, the existence of
such solutions can be seen from Eq. (\ref{linear fan}).
They are given by a difference of the amplitudes
depending on transverse coordinates of quark $\rho _{1}$
and antiquark $\rho _{2}$, respectively. Splitting of the
dipole does not lead to any evolution of such a solution:
new quarks and antiquarks on the edges of the dipole have
the same transverse coordinates, and therefore their
contributions cancel in the linear approximation.

\section{Solutions depending only on the dipole size }

Next let us investigate the simple case when the solution
does not depend on $b$, which, formally, corresponds  to
the asymptotics of $s(b,p,y)$ at $b\rightarrow \infty $
\begin{equation}
\lim_{b\rightarrow \infty }\,s(b,p,y)=s(p,y).
\end{equation}
This function describes the $S$-matrix for the scattering
of the dipole with the size $\rho $ on the hadron system
with a large radius $R\gg \rho $, provided that the impact
parameter $b$ is restricted to the interval
\[
R\gg b\gg \rho ~.
\]
In this case the fan equation is significantly
simplified. In terms of variables
\begin{equation}
\xi =\frac{1}{2}%
(y-p)\,, \;\;\; \eta =\frac{1}{2}(y+p)   \,\label{ksi,
eta}
\end{equation}
it takes the form:
\begin{equation}
\frac{\partial }{\partial \xi }\,s(\xi, \eta)=s^{2}(\xi,
\eta)\,.
\end{equation}
In accordance with the fact that this is a degenerated
Riccati equation, this equation has the
Kowalewskaya-Painleve property \cite{KovPen}, and its
solution is a simple meromorphic function
\begin{equation}
s(\xi, \eta)=\frac{1}{\xi (\eta )-\xi }\,.
\end{equation}
Returning to the $\rho $-representation, we obtain the
following result
\begin{equation}
S(\rho ,y)=\int_{-i\infty }^{i\infty }\frac{dp}{2\pi i}\,e^{\rho p}\,\frac{1%
}{\left(s(p+y)\right)^{-1}-y}\,,\,
\end{equation}
where $s(p)$ is the Laplace transform of the $S-$matrix
at the initial moment $y=0$%
\begin{equation}
s(p)=\int_{0}^{\infty }d\rho \,e^{-\rho p}\,S(\rho ,0)\,.
\end{equation}

Assuming the colour transparency property in the form
\begin{equation}
\lim_{\rho \rightarrow 0}S(\rho ,0)=1-c\rho ^{\gamma
}+...\,,\,\,\,\gamma
>0\,,\,c>0\,,
\end{equation}
one can calculate the function $s(p,y)$ at large $y$ or
$p$
\begin{equation}
s(p,y)\simeq \frac{1}{p+c\,\Gamma (1+\gamma
)\,(p+y)^{1-\gamma }}~.
\end{equation}
Therefore, if $0<\gamma <1$,  we obtain in the region
$p\sim y^{1-\gamma }\ll y\,$\ the scaling behaviour
\begin{equation}
s(p,y)\simeq \frac{1}{p+c\,\Gamma (1+\gamma
)\,y^{1-\gamma }}\,,
\end{equation}
which leads to the $S$-matrix
\begin{equation}
S(\rho ,y)\simeq e^{-c\Gamma (1+\gamma )\,y^{1-\gamma
}\rho }
\end{equation}
in the region $y^{1-\gamma }\rho \sim 1$. It means, that
at large $y$ the
distribution in $\rho $ is proportional to a smeared $\delta $-function $%
\delta (\rho ).$

If, on the other hand, $\gamma >1$ (for example, when
$\lim_{\rho \rightarrow 0}S(\rho ,0)=1-c\rho ^{2}$) the
$S$-matrix tends to unity in the large region $\rho \ll
y^{\gamma -1}$, which corresponds to the vanishing of the dipole
interaction. In the particular case when $\gamma =1$ and
\begin{equation}
\lim_{\rho \rightarrow 0}S(\rho ,0)=1-c\rho
+...\,,\,\,\,c>0\,,
\end{equation}
the asymptotic behaviour of the $S$-matrix at large $y$
is especially simple:
\begin{equation}
\lim_{y\rightarrow \infty }S(\rho ,y)=e^{-c\rho }\,.
\end{equation}
It corresponds to the fact that in the $2+1$-dimensional
QCD the fan equation has the stationary solutions
\begin{equation}
S(\rho ,y)=e^{-c\rho }\,,
\end{equation}
corresponding to the conservation of the total length of
dipoles. This expression is similar to the Maxwell
distribution in energies for molecules in a thermal
equilibrium. In our case the energy is proportional to the
total length $\rho$ of dipoles. Thus, there is an analogy between
the Boltzman and BK equations.

\section{Stationary solutions of the fan equation}

The stationary fan equation
\begin{equation}
0=\int_{\rho _{1}}^{\rho _{2}}d\rho _{0}\left( S_{\rho
_{2}\rho _{0}}\,S_{\rho _{0}\rho _{1}}-S_{\rho _{2}\rho
_{1}}\right) \,
\end{equation}
has the following solution
\begin{equation}
\widetilde{S}_{\rho _{2}\rho _{1}}=\frac{S(\rho
_{2})}{S(\rho _{1})}\,,
\end{equation}
where $S(\rho )$ is an arbitrary function. In a linear
approximation it corresponds to the amplitude $N(b,\rho
,y)$  (\ref{stationary N}) for the BFKL equation, being
the sum of two functions depending on $\rho _{1}$ and
$\rho _{2}$. In order to show that $\widetilde{S}_{\rho _{2}\rho
_{1}}$ is a general solution, we expand the equation in terms of
small fluctuations $\Delta S_{\rho _{2}\rho _{1}}$ around
the solution
\begin{equation}
\Delta S_{\rho _{2}\rho _{1}}=S_{\rho _{2}\rho
_{1}}-\widetilde{S}_{\rho _{2}\rho _{1}},
\end{equation}
and we obtain, in the linear approximation,
\begin{equation}
0=\int_{\rho _{1}}^{\rho _{2}}d\rho _{0}\left(
\widetilde{S}_{\rho _{2}\rho
_{0}}\,\Delta S_{\rho _{0}\rho _{1}}+\Delta S_{\rho _{2}\rho _{0}}\,%
\widetilde{S}_{\rho _{0}\rho _{1}}-\Delta S_{\rho
_{2}\rho _{1}}\right) \,.
\end{equation}
By introducing the relative correction $\widetilde{\delta
S}_{\rho _{2}\rho _{1}}$
\begin{equation}
\Delta S_{\rho _{2}\rho _{1}}=\widetilde{S}_{\rho _{2}\rho _{1}}\widetilde{%
\delta S}_{\rho _{2}\rho _{1}}\,,
\end{equation}
we simplify the above equation
\begin{equation}
0=\int_{\rho _{1}}^{\rho _{2}}d\rho _{0}\left(
\widetilde{\delta S}_{\rho
_{0}\rho _{1}}+\widetilde{\delta S}_{\rho _{2}\rho _{0}}-\widetilde{\delta S}%
_{\rho _{2}\rho _{1}}\right) \,.
\end{equation}
There is an obvious solution of this equation
\begin{equation}
\widetilde{\delta S}_{\rho _{2}\rho _{1}}=s(\rho
_{2})-s(\rho _{1}),
\end{equation}
where $s(\rho )$ is an arbitrary function. This correction, however, 
represents only a redefinition of the function $S(\rho )$ in the 
expression for $\widetilde{S}%
_{\rho _{2}\rho _{1}}$. In fact, by introducing the new
variables $\rho =\rho _{2}-\rho _{1}$, $b=(\rho _{2}+\rho
_{1})/2$ we can reduce the above
equation for $\delta S_{\rho _{2}\rho _{1}}$ to the stationary equation for $%
N(b,\rho ,y)$, which has only the solutions
$\widetilde{\delta S}_{\rho _{2}\rho _{1}}$, as it was
shown in Section 3.

The stationary fan equation has the Kowalewskaya-Painleve
property. Namely, if we the solution in the series
\begin{equation}
S_{\rho _{2}\rho _{1}}=\sum_{n=0}^{\infty }a_{n}(b)\,\rho
^{n};\;\;\;a_{0}=1,
\end{equation}
the first coefficient $a_{1}(b)$ is an arbitrary
function of $\rho $ with the following asymptotic
behaviour at $b\rightarrow \infty $
\begin{equation}
a_{1}(b)=\sum_{r=1}^{\infty }\frac{c_{r}}{b^{r}}\,.
\end{equation}
Indeed, the above solution of the stationary equation can
be written as follows
\begin{equation}
\widetilde{S}_{\rho _{2}\rho _{1}}=\prod_{k=1}^{\infty }\left( \frac{1+\frac{%
\rho }{2(b+d_{k})}}{1-\frac{\rho }{2(b+d_{k})}}\right)
^{r_{k}}\,\,,\,
\end{equation}
where $d_{k}$ and $r_{k}$\ are arbitrary complex
parameters.\thinspace Therefore, the above function
$a_{1}(b)$ is
\begin{equation}
a_{1}(b)=\sum_{k=1}^{\infty }\frac{r_{k}}{b+d_{k}}\,.
\end{equation}
In contrast to this, as a consequence of the Kowalewskaya-Penleve property, 
for a given $a_{1}(b)$ the solution $\widetilde{S}_{\rho _{2}\rho
_{1}}$\thinspace is obtained in an unique way. For
example, if we start from the scaling ansatz
\begin{equation}
a_{1}(b)=\frac{a}{b}\,,
\end{equation}
the stationary solution is
\begin{equation}
\widetilde{S}_{\rho _{2}\rho _{1}}=\left( \frac{b+\frac{\rho }{2}}{b-\frac{%
\rho }{2}}\right) ^{a}\simeq 1+\,a\,\frac{\rho }{b}+\frac{a^{2}}{2}\frac{%
\rho ^{2}}{b^{2}}+\left( \frac{a}{12}+\frac{a^{3}}{6}\right) \frac{\rho ^{3}%
}{b^{3}}+O\left( \rho ^{4}\right) ~.\label{simple
solution}
\end{equation}
We can also write solutions of the stationary equation which
do not have any singularities at real $\rho _{1,2}$. For example,
\begin{equation}
\widetilde{S}_{\rho _{2}\rho _{1}}=\frac{1+\rho
_{1}^{2}}{1+\rho _{2}^{2}}\,.
\end{equation}

\section{Scaling solution of the non-stationary equation}

Let us now consider the general case when $s(b,p,y)$ depends
on $b$, and let us introduce the new variables
\begin{equation}
\,\eta =\frac{1}{2}(y+p)\,,\,\,\epsilon =\xi (\eta
)-\frac{1}{2}(y-p)\,,\,\,
\end{equation}
where $\xi (\eta )$ is an arbitrary function which later on will
be derived from the requirement that $s(b,p,y)$ satisfies initial 
conditions at $y=0$. In terms of the new variables the fan equation 
looks as follows
\[
-\frac{\partial }{\partial \epsilon }\,S(b,\epsilon ,\eta
)=\lim_{\eta _{r}\rightarrow \eta ,\epsilon
_{r}\rightarrow \epsilon }\overline{N}\,
\]
\begin{equation}
\left( S(b+\frac{1}{4}\left( \frac{\partial }{\partial
\eta _{2}}+\left(
1+\xi ^{\prime }(\eta _{2})\right) \frac{\partial }{\partial \epsilon _{2}}%
\right) ,\epsilon _{1},\eta _{1})\,S(b-\frac{1}{4}\left( \frac{\partial }{%
\partial \eta _{1}}+\left( 1+\xi ^{\prime }(\eta _{1})\right) \frac{\partial
}{\partial \epsilon _{1}}\right) ,\epsilon _{2},\eta
_{2})\right) 1\,.\, \label{differential b-dependent
equation}
\end{equation}
Its solution can be searched in the form of an expansion in powers of 
$1/\epsilon$:
\begin{equation}
\widetilde{S}(b,\epsilon ,\eta )=\frac{1}{\epsilon
}+\sum_{r=1}^{\infty }\Delta _{r}\widetilde{S}(b,\xi
,\eta )\,,\,\Delta _{r}\widetilde{S}(b,\xi ,\eta
)=\frac{c_{r}(b,\eta )}{\epsilon ^{r+1}}\,.
\label{Painleve form}
\end{equation}
Here $c_{1}(b,\eta )$ turns out to be an arbitrary
function, fixed by initial
conditions for $s(b,p,y)$ at $y=0$. According to the fan equation the 
residues $c_{r}(b,\eta )$, near $b=\infty$, 
are analytic functions with the following asymptotic behaviour:
\begin{equation}
c_{r}(b,\eta )=\frac{1}{b^{r}}\sum_{s=0}^{\infty }\frac{c_{r,s}(\eta )}{b^{s}}
\,.
\end{equation}
In particular, for the first correction $\Delta
_{1}S(b,\epsilon ,\eta )$ to the pole $1/\epsilon $ we
obtain the differential equation
\begin{equation}
\left( -\frac{\partial }{\partial \epsilon
}-\frac{2}{\epsilon }\right) \Delta _{1}S(b,\epsilon
,\eta )=0\,.
\end{equation}
Its solution is
\begin{equation}
\Delta _{1}S(b,\epsilon ,\eta )=\frac{c_{1}(b,\eta
)}{\epsilon ^{2}}\,,
\end{equation}
where
\begin{equation}
c_{1}(b,\eta )=\frac{1}{b}\sum_{s=0}^{\infty
}\frac{c_{1,s}(\eta )}{b^{s}}\,
\end{equation}
is an arbitrary function which is analytic and vanishing
at $b\rightarrow
\infty $. It should be chosen to satisfy the initial conditions for $%
s(b,\rho ,y)$ at $y=0$.

Therefore, in the expansion of $S(b,\epsilon ,\eta )$ in powers of 
$1/\epsilon, $ there is a phenomenon similar to the
Kowalewskaya-Painleve property for the integrable
differential equations. Namely, one can construct the
general solution in the class of meromorphic functions by
fixing the functions $\xi (\eta )$ and $c_{1}(b,\eta )$
from the initial conditions for $s(b,p,y)$. The next
correction $\Delta _{2}S(b,\epsilon ,\eta )$ satisfies
the equation
\[
\left( -\frac{\partial }{\partial \epsilon
}-\frac{2}{\epsilon }\right) \,\Delta _{2}\,S(b,\epsilon
,\eta )=
\]
\begin{equation}
\frac{1}{\epsilon ^{4}}\left( \overline{N}\,c_{1}(b+\frac{1}{4}\frac{%
\partial }{\partial \eta _{2}},\eta _{1}).c_{1}(b-\frac{1}{4}\frac{\partial
}{\partial \eta _{1}},\eta _{2})_{|_{\eta _{2}=\eta _{1}=\eta }}-\frac{1}{2}%
\frac{\partial }{\partial
b}c_{1}(b+\frac{1}{4}\frac{\partial }{\partial \eta
_{2}},\eta )\left( 1+\xi ^{\prime }(\eta _{2})\right)
_{|_{\eta _{2}=\eta }}\right) ,
\end{equation}
where, in the last term, it is implied that after the differentiation of $%
c_{1}$ the removed operator $b+\frac{1}{4}\frac{\partial }{\partial \eta _{2}%
}$ is substituted by the expression $1+\xi ^{\prime }(\eta
_{2})$ in the corresponding place. Therefore we obtain
\begin{equation}
\Delta _{2}S(b,\epsilon ,\eta )=\frac{c_{2}(b,\eta
)}{\epsilon ^{3}}\,,
\end{equation}
where
\begin{equation}
c_{2}(b,\eta
)=\overline{N}\,c_{1}(b+\frac{1}{4}\frac{\partial
}{\partial
\eta _{2}},\eta _{1}).c_{1}(b-\frac{1}{4}\frac{\partial }{\partial \eta _{1}}%
,\eta _{2})_{|_{\eta _{2}=\eta _{1}=\eta }}-\frac{1}{2}\frac{\partial }{%
\partial b}c_{1}(b+\frac{1}{4}\frac{\partial }{\partial \eta _{2}},\eta
)\left( 1+\xi ^{\prime }(\eta _{2})\right) _{|_{\eta
_{2}=\eta }}\,.
\end{equation}
In an analogous way one can calculate the residues
$c_{r}(b,\eta )$ in terms of $\xi (\eta )$ and
$c_{k}(b,\eta )$ with $k<r$.

Let us now consider the solution of the fan equation in
the scaling regime
\begin{equation}
\epsilon \ll 1\,,\,\,b\gg 1\,,\,\,\epsilon b\sim 1,
\end{equation}
for fixed $\eta $. In this case the equation
(\ref{differential b-dependent equation}) is
significantly simplified
\begin{equation}
-\frac{\partial }{\partial \epsilon }\,S(b,\epsilon
)=\lim_{\epsilon
_{r}\rightarrow \epsilon }\overline{N}\,\left( S(b+u\frac{\partial }{%
\partial \epsilon _{2}},\epsilon _{1})\,S(b-u\frac{\partial }{\partial
\epsilon _{1}},\epsilon _{2})\right) 1\,, \label{DL
equation}
\end{equation}
where the parameter $u=u(\eta )$ is defined by the
expression
\begin{equation}
u=\frac{1}{4}\left( 1+\xi ^{\prime }(\eta )\right) \,,
\end{equation}
and the solution has the following expansion at large
$\epsilon b$
\begin{equation}
\,S(b,\epsilon )=\frac{1}{\epsilon }+\frac{v}{\epsilon
^{2}b}+...\,. \label{S(b,epsilon )}
\end{equation}
Here the function $v=v(\eta )$ is fixed by the initial conditions for $%
s(b,\,p,\,y)$. The first two terms on the rhs of (\ref{S(b,epsilon )})
determine all the other coefficients of the expansion
\begin{equation}
S(b,\epsilon )=\frac{1}{\epsilon }\sum_{n=0}^{\infty
}\frac{a_{n}}{(\epsilon
b)^{n}};\;\;\;a_{0}=1,\;\;a_{1}=v.  \label{expansion
S(b,epsilon)}
\end{equation}
The reccurence relations for the coefficients $a_{n}$ are
given in the Appendix, and one obtains
\begin{equation}
S(b,\epsilon )=\frac{1}{\epsilon }+\frac{v}{\epsilon ^{2}b}+\frac{v^{2}}{%
\epsilon ^{3}b^{2}}+\left( 2u^2v+v^{3}\right)
\frac{1}{\epsilon ^{4}b^{3}}+...
\end{equation}

It is easy to see that the substitution
\begin{equation}
S(b,\epsilon)\rightarrow \frac{1}{2u}
s(b,\frac{\epsilon}{2u})\,, \;\; \epsilon \rightarrow
{2u}{p}
\end{equation}
turns (\ref{DL equation}) into (\ref{equation in mixed
representation}) for the stationary ($y$-independent)
case. Therefore we have the remarkable result: the
general solution of the non-stationary equation at
$\epsilon b\sim 1$ can be expressed in terms of the
stationary solution depending on $\rho /b$. Using the
stationary solution obtained in the previous section (see
(\ref{simple solution})) one can derive
\begin{equation}
S(b,\epsilon )=\frac{1}{2u}\int_{0}^{\infty }d\rho \,
e^{-\frac{\epsilon \rho}{2u} }\left( \frac{b+%
\frac{\rho }{2}}{b-\frac{\rho }{2}}\right)
^{\frac{v}{2u}}\,.
\end{equation}

\section{Infinite parameter symmetry of the fan equation}

In this section we consider another ansatz:
\begin{equation}
{S}_{\rho _{2}\rho _{1}}(y)=\widetilde{S}_{\rho _{2}\rho
_{1}}(y)=e^{(\chi (\rho _{2})-\chi (\rho
_{1}))y}\,\frac{R(\rho _{2})}{T(\rho _{1})}\,,
\end{equation}
where $R(\rho )$ and $T(\rho )$ are arbitrary functions.
Putting this ansatz into Eq.(\ref{fan}) it is easy to
verify that it is indeed a non-stationary solution,
provided that
\begin{equation}
\chi (\rho )=\int_{0}^{\rho }d\rho _{0}\left(
\frac{R(\rho _{0})}{T(\rho _{0})}-1\right) \,.
\end{equation}
For small fluctuations $\delta N_{\rho _{2}\rho _{1}}(y)$
around this solution
\begin{equation}
S_{\rho _{2}\rho _{1}}(y)\simeq \widetilde{S}_{\rho
_{2}\rho _{1}}(y)(1+\delta N_{\rho _{2}\rho _{1}}(y))\,
\label{fluctuations}
\end{equation}
we obtain, inserting (\ref{fluctuations}) into (\ref{fan})
and linearizing the result in $\delta N_{\rho _{2}\rho
_{1}}(y))$,
\begin{equation}
\frac{\partial \delta N_{\rho _{2}\rho _{1}}(y)}{\partial
y}=\int_{\rho _{1}}^{\rho _{2}}d\rho _{0}\frac{R(\rho
_{0})}{T(\rho _{0})}\left( \delta N_{\rho _{2}\rho
_{0}}(y)+\delta N_{\rho _{0}\rho _{1}}(y)-\delta N_{\rho
_{2}\rho _{1}}(y)\right) \,.
\end{equation}
We have ''zero mode'' solutions of this equation
\begin{equation}
\widetilde{\delta N}_{\rho _{2}\rho _{1}}(y)=\delta
R(\rho _{2})-\delta T(\rho _{1})+y\left( \delta \chi
(\rho _{2})-\delta \chi (\rho _{1})\right) \,,
\end{equation}
where $\delta R(\rho )$ and $\delta T(\rho )$ are
arbitrary functions related to $\delta \chi (\rho )$ as
follows
\begin{equation}
\delta \chi (\rho )=\int_{0}^{\rho }d\rho
_{0}\frac{R(\rho _{0})}{T(\rho _{0})}\left( \delta R(\rho
_{0})-\delta T(\rho _{0})\right) \,.
\end{equation}

Introducing the new integration variable $\Upsilon(\rho)$
\begin{equation}
\Upsilon(\rho) =\int^{\rho}d\rho _{0}\frac{R(\rho
_{0})}{T(\rho _{0})},\label{Upsilon}
\end{equation}
we obtain the following linear equation for $\delta
N_{\rho _{2}\rho _{1}}(y)\equiv N_{\Upsilon _{2}\Upsilon
_{1}}(y)$:
\begin{equation}
\frac{\partial N_{\Upsilon _{2}\Upsilon
_{1}}(y)}{\partial y}=\int_{\Upsilon _{1}}^{\Upsilon
_{2}}d\Upsilon_{0}\,\left( N_{\Upsilon _{2}\Upsilon
_{0}}(y)+N_{\Upsilon _{0}\Upsilon _{1}}(y)-N_{\Upsilon
_{2}\Upsilon _{1}}(y)\right) \,,
\end{equation}
which coincides with the BFKL equation having the general
solution given by (\ref{solution of linear fan}) with the
substitution
\[
\rho =\rho _{21}\rightarrow \Upsilon_{2}-\Upsilon_{1}\,,\;\;b
\rightarrow \frac{\Upsilon_{2}+\Upsilon_{1}}{2}\,; \;\;\;
\Upsilon_{r}\equiv \Upsilon(\rho_{r})\,.
\]

Let us factorize $S_{\rho _{2}\rho _{1}}(y)$ as follows
\begin{equation}
S_{\rho _{2}\rho _{1}}(y)\simeq \widetilde{S}_{\rho
_{2}\rho _{1}}(y)\,s_{\rho _{2}\rho _{1}}(y)\,.
\end{equation}
Then we obtain the following equation for $s_{\rho _{2}\rho _{1}}(y))$%
\begin{equation}
\frac{\partial s_{\rho _{2}\rho _{1}}(y)}{\partial
y}=\int_{\rho _{1}}^{\rho _{2}}d\rho _{0}\frac{R(\rho
_{0})}{T(\rho _{0})}\left( s_{\rho _{2}\rho
_{0}}(y)s_{\rho _{0}\rho _{1}}(y)-s_{\rho _{2}\rho
_{1}}(y)\right)\,.
\end{equation}

Therefore the fan equation has a renormalization group property. Namely, $%
s_{\rho _{2}\rho _{1}}(y)$ satisfies the same equation
as $S_{\rho _{2}\rho _{1}}(y)$, but in the new variables
$\Upsilon $, defined by (\ref{Upsilon}). As a result, we
obtain the following ''Backlund'' transformation between
different solutions of the fan equation
\begin{equation}
e^{(\rho _{2}-\rho _{1})\,y}S_{\rho _{2}\rho _{1}}^{(1)}(y)=\,\,\frac{%
R_{12}(\rho _{2})}{T_{12}(\rho _{1})}\,e^{(\Upsilon
_{2}-\Upsilon _{1})\,y}\,S_{\Upsilon _{2}\Upsilon
_{1}}^{(2)}(y)
\end{equation}
for arbitrary $R(\rho )$ and $T(\rho )$, provided that the new variable 
satisfies
\begin{equation}
\Upsilon (\rho )=\int^{\rho }d\rho _{0}\frac{R_{12}(\rho
_{0})}{T_{12}(\rho _{0})}\,. \label{Upsilon-rho}
\end{equation}
The function $\Upsilon (\rho)$ is chosen to grow with
$\rho$ and to have the additional constraint
\begin{equation}
\Upsilon (\pm \infty)=\pm \infty \,.
\end{equation}

We can once more perform the above transformation:
\begin{equation} e^{(\Upsilon _{2}-\Upsilon
_{1})\,y}\,S_{\Upsilon _{2}\Upsilon
_{1}}^{(2)}(y)=\,\,\frac{R_{23}(\Upsilon _{2})}{T_{23}(\Upsilon
_{1})} e^{(\Gamma _{2}-\Gamma _{1})\,y}\,S_{\Gamma
_{2}\Gamma _{1}}^{(3)}(y)\,,\,\,\Gamma _{r}=\Gamma
(\Upsilon _{r})\,,
\end{equation}
where
\begin{equation}
\Gamma (\Upsilon )=\int^{\Upsilon }d\Upsilon
_{0}\frac{R_{23}(\Upsilon _{0})}{T_{23}(\Upsilon
_{0})}\,,
\end{equation}
and verify its group property
\begin{equation}
e^{(\rho _{2}-\rho _{1})\,y}S_{\rho _{2}\rho _{1}}^{(1)}(y)=\,\,\frac{%
R_{13}(\rho _{2})}{T_{13}(\rho _{1})}\,e^{(\Gamma
_{2}-\Gamma _{1})\,y}\,S_{\Gamma _{2}\Gamma
_{1}}^{(3)}(y)\,,\,\,\Gamma _{r}=\Gamma (\Upsilon
(\rho _{r}))\,.
\end{equation}
Here
\begin{equation}
\frac{R_{13}(\rho _{2})}{T_{13}(\rho _{1})}=\frac{R_{12}(\rho _{2})}{%
T_{12}(\rho _{1})}\,\frac{R_{23}(\Upsilon
_{2})}{T_{23}(\Upsilon _{1})}
\end{equation}
and
\begin{equation}
\Gamma (\Upsilon (\rho ))=\int^{\Upsilon (\rho )}d\Upsilon
_{0}\frac{ R_{23}(\Upsilon _{0})}{T_{23}(\Upsilon
_{0})}=\int^{\rho }d\rho _{0}\frac{ R_{12}(\rho
_{0})}{T_{12}(\rho _{0})}\,\frac{R_{23}(\Upsilon (\rho _{0}))}{%
T_{23}(\Upsilon (\rho _{0}))}\,.
\end{equation}
In particular, for
\begin{equation}
\Gamma (\Upsilon (\rho ))=\rho \,
\end{equation}
one obtains
\begin{equation}
\frac{R_{12}(\rho )}{T_{12}(\rho )}\,\frac{R_{23}(\Upsilon (\rho ))}{%
T_{23}(\Upsilon (\rho ))}=1\,.
\end{equation}
Generally,
\begin{equation}
\frac{R_{13}(\rho _{2})}{T_{13}(\rho _{1})}=\frac{R_{12}(\rho _{2})}{%
T_{12}(\rho _{1})}\,\frac{R_{23}(\Upsilon
_{2})}{T_{23}(\Upsilon _{1})}\neq 1\,,
\end{equation}
and we thus obtain a family of
solutions
\begin{equation}
S_{\rho _{2}\rho _{1}}^{(f)}(y)=\,\,\frac{f(\rho _{2})}{f(\rho _{1})}%
\,S_{\rho _{2}\rho _{1}}(y)\,
\end{equation}
for arbitrary $f(\rho )$. The symmetry group of the fan
equation is a product of this abelian transformation of
the dipole $S$-matrix $S_{\rho _{2}\rho _{1}}(y)$\ and
of the reparametrization group (\ref{Upsilon-rho}).

Without any loss of generality it is convenient to
write the action of the second group on $S_{\rho _{2}\rho
_{1}}(y)$ in the form
\begin{equation}
e^{(\rho _{2}-\rho _{1})\,y}S_{\rho _{2}\rho
_{1}}^{(a)}(y)=\,a(\rho _{2})\,a(\rho _{1})\,e^{(\Upsilon
_{2}-\Upsilon _{1})\,y}\,S_{\Upsilon _{2}\Upsilon
_{1}}(y)\,,\,\,\Upsilon _{r}=\Upsilon (\rho
_{r})=\int^{\rho_r}d\rho \,a^{2}(\rho )\,,
\end{equation}
where $a(\rho )=\sqrt{R_{12}(\rho )/T_{12}(\rho )}$. We
can express this equation directly in terms of the
transformation $\rho ^{\prime }=\Upsilon (\rho )$
\begin{equation}
e^{(\rho _{2}-\rho _{1})\,y}S_{\rho _{2}\rho _{1}}^{(1)}(y)=\,\sqrt{\frac{%
d\Upsilon (\rho _{2})}{d\rho _{2}}\frac{d\Upsilon (\rho _{1})}{d\rho _{1}}}%
\,e^{(\Upsilon _{2}-\Upsilon _{1})\,y}\,S_{\Upsilon
_{2}\Upsilon _{1}}^{(2)}(y)\,.
\end{equation}
It means that under the general covariance
transformation $\rho \rightarrow \Upsilon (\rho )$the quantity
\begin{equation}
T_{\rho _{2}\rho _{1}}(y)=e^{(\rho _{2}-\rho
_{1})y}\,S_{\rho _{2}\rho _{1}}(y)\,
\end{equation}
transforms as a
tensor with the spinor indices $\rho _{2},\rho _{1}$. The
functional expressing the $S$-matrix $S_{\rho _{2}\rho
_{1}}(y)$ through initial conditions
\begin{equation}
S_{\rho _{2}\rho _{1}}(y)=\Theta \left( S_{\rho _{2}\rho
_{1}}(0)\right)
\end{equation}
should be invariant under the full symmetry group
(together with the above multiplication of $S_{\rho
_{2}\rho _{1}}^{(1)}$ by the ratio $f(\rho _{2})/f(\rho
_{1})$). In particular we can use the above
transformations to change in a convenient way the initial
conditions
\begin{equation}
S_{\rho _{2}\rho _{1}}^{(1)}(0)=\,\frac{R(\rho _{2})}{T(\rho _{1})}%
\,S_{\Upsilon _{2}\Upsilon _{1}}^{(2)}(0)\,.
\end{equation}

\section{General solution of the fan equation}

The fan equation for the function
\begin{equation}
G_{\rho _{2}\rho _{1}}(y)=e^{(\rho _{2}-\rho
_{1})y}\,S_{\rho _{2}\rho _{1}}(y)\,
\end{equation}
can be written as follows
\begin{equation}
\frac{\partial G_{\rho _{2}\rho _{1}}(y)}{\partial
y}=\int_{\rho _{1}}^{\rho _{2}}d\rho _{0}\,G_{\rho
_{2}\rho _{0}}(y)\,G_{\rho _{0}\rho _{1}}(y)\,.
\end{equation}
The function $G_{\rho _{2}\rho _{1}}(y)$ can be
considered as a matrix element of the
infinite-dimensional matrix \ $G(u)$ having a triangular
form, in accordance with the fact that it is nonzero
only for $\rho _{2}\geq \rho _{1}$
\begin{equation}
G_{\rho _{2}\rho _{1}}(y)=\langle\rho _{2}|G|\rho
_{1}\rangle\,.
\end{equation}
We can define the product of two matrices $G^{a}(u)$ and
$G^{b}(u)$ as the matrix $G^{c}(u)$ with the matrix
element
\begin{equation}
G_{\rho _{2}\rho _{1}}^{c}(y)=\int_{\rho _{1}}^{\rho
_{2}}d\rho _{0}\,G_{\rho _{2}\rho _{0}}^{a}(y)\,G_{\rho
_{0}\rho _{1}}^{b}(y)\,.
\end{equation}
Correspondingly, the inverse matrix $G^{-1}(u)$ is defined as the matrix
satisfying the relations
\begin{equation}
\delta (\rho _{2}-\rho _{1})=\int_{\rho _{1}}^{\rho
_{2}}d\rho _{0}\,G_{\rho _{2}\rho
_{0}}(y)\,(G^{-1})_{\rho _{0}\rho _{1}}(y)=\int_{\rho
_{1}}^{\rho _{2}}d\rho _{0}\,(G^{-1})_{\rho _{2}\rho
_{0}}(y)\,G_{\rho _{0}\rho _{1}}(y)\,.
\end{equation}
Its matrixelements $\,(G^{-1})_{\rho _{2}\rho _{1}}(y)$
are non-zero only for  $\rho _{2}\geq \rho _{1}$. Thus,
the operators $G(y)$ belong to the (solvable) group of
triangular matrices. The matrix element of the unit
element of this group is $\delta (\rho _{2}-\rho _{1})$.

In the operator form the fan equation looks as follows
\begin{equation}
\frac{\partial G(y)}{\partial y}=G^{2}(y)\,,
\end{equation}
and its general solution has the meromorphic property of
Kowalewskaya-Painleve \cite{KovPen}
\begin{equation}
G(y)\,=\frac{G(0)}{1-y\,G(0)}\,\,.
\end{equation}
Note that in accordance with the Ricatti theory
the operator $G^{-1}$ satisfies the free Newton equaton
\begin{equation}
\frac{d^2}{dy^2}\,G^{-1}(y)=0.
\end{equation}
The corresponding solution for the $S$-matrix $S_{\rho
_{2}\rho _{1}}(y)$ written in the operator form is
\begin{equation}
S(y)\,=e^{-\rho y}\frac{G(0)}{1-y\,G(0)}\,\,e^{\rho y},
\end{equation}
where $\rho $ is the operator with the eigenvalues $\rho
_{1}$ for the eigenfunctions $|$$\rho _{1}\rangle$. This
solution is similar to the general solution of a linear
equation expressed in terms of its Green function.
In a matrix form it can
be expanded in a series  as follows
\[
e^{(\rho _{2}-\rho _{1})\,y}\,S_{\rho _{2}\rho
_{1}}(y)=S_{\rho _{2}\rho _{1}}(0)+
\]
\begin{equation}
y\int_{\rho _{1}}^{\rho _{2}}d\rho _{0}\,S_{\rho _{2}\rho
_{0}}(0)\,S_{\rho _{0}\rho _{1}}(0)+y^{2}\,\int_{\rho
_{1}}^{\rho _{2}}d\rho _{0^{\prime }}\,\int_{\rho
_{1}}^{\rho _{0^{\prime }}}d\rho _{0}\,S_{\rho _{2}\rho
_{0^{\prime }}}(0)\,S_{\rho _{0^{\prime }}\rho
_{0}}(0)S_{\rho _{0}\rho _{1}}(0)+...\,.
\end{equation}
In particular, according to above discussions one can
simplify this general result as follows
\begin{equation}
\widetilde{S}_{\rho _{2}\rho _{1}}(y)=e^{(\chi (\rho
_{2})-\chi (\rho _{1}))y}\,\frac{R(\rho _{2})}{T(\rho
_{1})}\,,\,\,\chi (\rho
)=\int_{0}^{\rho }d\rho _{0}\left( \frac{R(\rho _{0})}{T(\rho _{0})}%
-1\right) ,
\end{equation}
providing that $S_{\rho _{2}\rho _{1}}(0)$ has the
factorized form
\begin{equation}
\widetilde{S}_{\rho _{2}\rho _{1}}(0)=\frac{R(\rho
_{2})}{T(\rho _{1})}\,.
\end{equation}
It is also obvious that the solutions $S_{\rho _{2}\rho
_{1}}(y)$ for different $S_{\rho _{2}\rho _{1}}(0)$ are
related each to other by the above established group
of transformations depending on two arbitrary functions
$f(\rho )$ and $\Upsilon (\rho )$
\begin{equation}
e^{(\rho _{2}-\rho _{1})\,y}S_{\rho _{2}\rho _{1}}^{(f,\Upsilon )}(y)
=\frac{f(\rho _{2})}{f(\rho _{1})}\,\sqrt{\frac{d\Upsilon
(\rho _{2})}{d\rho _{2}}\frac{d\Upsilon (\rho
_{1})}{d\rho _{1}}}e^{(\Upsilon _{2}-\Upsilon
_{1})\,y}\,S_{\Upsilon _{2}\Upsilon
_{1}}(y)\,,\,\,\Upsilon _{r}=\Upsilon (\rho _{r})\,.
\end{equation}

In particular, we can consider the case where the initial condition
$S_{\Upsilon _2 \Upsilon _1}(0)$ is only a function  of $\Upsilon
_{2}-\Upsilon _{1}$, which gives us a possibility to construct
the solution depending on three arbitrary functions
\begin{equation}
e^{(\rho _{2}-\rho _{1})\,y}S_{\rho _{2}\rho
_{1}}^{(f,\Upsilon ,s )} (y)=  \int_{\sigma -i\infty
}^{\sigma +i\infty }\frac{dp}{2\pi i}\, \frac{c^{(2)}_{\rho
_2}(p)\,c^{(1)}_{\rho _1}(p)}{(s(p))^{-1}-y}\,,
\end{equation}
where the parameter $\sigma$ is chosen from the condition that
all singularities of the integrand are situated to the left of
the integration contour in $p$ and
$$
s(p)=\int _0^{\infty}d\Upsilon \,e^{-\Upsilon
\,p}\,S_{\frac{\Upsilon}{2},-\frac{\Upsilon}{2}}(0)\,,
$$
\begin{equation}
c^{(2)}_{\rho
_2}(p)= \sqrt{\frac{d\Upsilon (\rho _{2})}{d\rho _2}}\,
e^{\Upsilon _2\,p}\,f(\rho _2)\,,\,\, c^{(1)}_{\rho _1}(p)=
\sqrt{\frac{d\Upsilon (\rho _{1})}{d\rho _1}}\,
\left(e^{\Upsilon _1\,p}f(\rho _1)\right)^{-1}\,.
\end{equation}
The solution $S^{(f,\Upsilon , s)}$ can be considered as
a particular case of
the general solution transformed to a diagonal
form by an appropriate similarity transformation. 
The function $s(p)$ has the interpretation of  the
eigenvalue of $S_{\rho _2 \rho _1}(0)$, and $c^{(2)}_{\rho
_1}(p)$ is its eigenfunction. The similarity
transformation is unitary
($c^{(2)}=c^{(1)\,+}=1/c^{(1)}$) provided that $f(\rho )$
is a phase: $f^*=f^{-1}$.

In general the functions $c^{(r)}_{\rho}(p)$ have a more
complicated (non-exponential) dependence on $p$ which
enumerates the eigenvalues. This dependence
 can be found from the solution
of the equation
\begin{equation}
\int _{-\infty}^{\infty} d\,\rho _1 \,S_{\rho _2\rho _1}(0)
c_{\rho _1}(p)=s(p)\,c_{\rho _2}(p)\,.
\end{equation}

For the triangular matrix $S_{\rho _2 \rho _1}(0)$ the
eigenfunction $c_{\rho }(p)$ can be calculated easily,
provided that we substitute the continuous coordinate $\rho$  by
a finite number of points $\rho _k$. In general, the behaviour of
$S_{\rho _2\rho _1}(y)$ at large $y$ depends on the initial
conditions encoded in the asymptotics of the functions
$c_{\rho}(p)$ at large $p$.  Due to the colour transparancy
property we have
\begin{equation}
S_{\rho _{2}\rho _{1}}(0)<1\,,\,\
\lim_{\rho _{2}\rightarrow \rho _{1}}S_{\rho _{2}\rho
_{1}}(0)\,=1-c\rho _{21}^{\,\gamma \,},\,\,\,\gamma >0\,\,,
\end{equation}
where the
coefficient $\,c>0$ is a function of $b$.
At large $y$ and fixed $\rho _{r}$ it is natural to expect
an universal dependence of $S_{\rho _2\rho _1}(y)$ from
$y$. In particular, for large $c$ the $S$-matrix should have the
form of a smeared $\delta (\rho _{21})$-function.
It is related to the fact that in this case, as a
result of the $y$-evolution, the average size of the dipole tends
to zero, and one can neglect the $b$-dependence of $S$ (up
to a common factor).  In the pre-asymptotics a scaling behaviour
of $S_{\rho _{2}\rho _{1}}(y)$ is possible (cf. section 4).

It is important that the solution of the Balitsky-Kovchegov equation
satisfies also the linear integral equation
\begin{equation}
G_{\rho _{2}\rho _{1}}(y)=G_{\rho _{2}\rho _{1}}(0)+y\,\int
^{\rho _2}_{\rho _1} d\,\rho \,\,G_{\rho _{2}\rho _{0}}(0)
\,G_{\rho _{0}\rho _{1}}(y)\,.
\end{equation}
In particular, the theory of the Fredholm equations can be applied to this 
equation. In this case, the perturbation theory in $y$ should work
rather well.  For large $y$ one can use a quasi-classical
approximation.  We hope to return to the investigation of the
this linear equation in future publications.

\section{Conclusions}
In this paper we have obtained analytic solutions of a simplified version of 
the nonlinear BK equation, namely the BK equation in one transverse dimension.
Our explicit formulae allow to study, for any initial condition, the 
rapidity evolution, both as a function of the dipole size and of the dipole 
position in impact parameter. 

We also have found several remarkable 
properties of the nonlinear equation: the general solution has the 
meromorphic Kowalewskaya-Painlev\'{e} property, and the equation has 
symmetries which are reminiscent of general covariance. Some of these 
results might be helpful in addressing also the more realistic case of two 
transverse dimensions.\\ \\     

\noindent     
{\bf Acknowledgements:}\\
This work was started while two of us (V.S.Fadin  and L.N. Lipatov) 
were visiting the II.Institut f\"ur Theoretische Physik, Universit\"at 
Hamburg, and DESY. We gratefully acknowledge their hospitality. 

\newpage
\noindent
{\Large \bf Appendix }

Here we derive the reccurence relations for the
coefficients $a_{n}$ in (\ref{expansion S(b,epsilon)}).
Inserting the expansion (\ref{expansion S(b,epsilon)}) in
(\ref{DL equation}%
) one can obtain the relation
$$
(n+1)\frac{1}{\epsilon ^{2}}\sum_{n=0}^{\infty }\frac{a_{n}}{(\epsilon b)^{n}%
}=\frac{1}{\epsilon ^{2}}\sum_{k=0}^{\infty }\sum_{l=0}^{\infty }\frac{a_{k}%
}{(\epsilon b)^{k}}\frac{a_{l}}{(\epsilon b)^{l}}\left(
(1+\frac{u}{\epsilon
b}\frac{d}{dx})^{-k}\frac{1}{x^{l+1}}\right) |_{x=1}\left( (1-\frac{u}{%
\epsilon b}\frac{d}{dx})^{-l}\frac{1}{x^{k+1}}\right)
|_{x=1}~.
$$
Using the expansion
$$
\left(
\frac{1}{(1+z\frac{d}{dx})^{k}}\frac{1}{x^{l+1}}\right)
|_{x=1}=\sum_{r=0}^{\infty }\frac{\Gamma (k+r)}{\Gamma (r+1)\Gamma (k)}%
\left( \left( -z\frac{d}{dx}\right)
^{r}\frac{1}{x^{l+1}}\right) |_{x=1}=\sum_{r=0}^{\infty
}C_{k,l+1}^{r}z^{r}~,
$$
where
$$
C_{k,l}^{r}=\frac{\Gamma (k+r)\Gamma (l+r)}{\Gamma (r+1)\Gamma (k)\Gamma (l)}%
~,
$$
one can easily verify that $a_{0}=1$, and $a_{1}$ is arbitrary.
For the coefficients $a_n$ with  $n>1$ we obtain
the recurrence relation
$$
(n-1)a_n=
$$
$$
\sum_{k=1}^{n-1}a_k\,C^{n-k}_{k,1}\left(u^{n-k}+
(-u)^{n-k}\right)
+\sum_{m=2}^{n}\sum_{k=1}^{m-1}a_k\;a_{m-k}\,u^{n-m}\;
\sum_{l=0}^{n-m}
(-1)^l\,C^{l}_{k,m-k+1}\;C^{n-m-l}_{m-k,k+1}~.
$$
Evidently, the terms with odd powers of $u$ are cancelled.

\end{document}